\documentclass[prb,longbibliography,showpacs, showkeys, twocolumn,superscriptaddress,amsmath,amssymb,verbatim]{revtex4-2}

\usepackage{graphicx}
\usepackage{lmodern}
\usepackage{epstopdf}
\usepackage{dcolumn}
\usepackage{bm}
\usepackage{subfigure}
\usepackage{amsmath} 
\usepackage{chemformula} 
\usepackage{siunitx} 

\usepackage[pdftex,colorlinks=true,citecolor=blue,linkcolor=blue,urlcolor=blue,bookmarks=true]{hyperref}
\usepackage[sort&compress]{natbib}

\usepackage{color}
\usepackage{textcomp}
\definecolor{red}{rgb}{1,0,0}

\definecolor{blue}{rgb}{0,0,1}

\definecolor{green}{rgb}{0,1,0}

\definecolor{lblue}{rgb}{0.4, 0.8, 0.8}

\begin{document}
\preprint{APS}

\def\scititle{
    Non-equilibrium Muon Dynamics Imitating Fractional Spin Excitations in a Quantum Spin Chain
}
\title{\scititle}

\author{Anton Hromov}
\affiliation{Jo\v{z}ef Stefan Institute, Jamova c. 39, 1000 Ljubljana, Slovenia}
\affiliation{Faculty of Mathematics and Physics, University of Ljubljana, Jadranska ulica 19, 1000 Ljubljana, Slovenia}
\author{Andrej Zorko}
\affiliation{Jo\v{z}ef Stefan Institute, Jamova c. 39, 1000 Ljubljana, Slovenia}
\affiliation{Faculty of Mathematics and Physics, University of Ljubljana, Jadranska ulica 19, 1000 Ljubljana, Slovenia}
\author{Matja\v{z} Gomil\v{s}ek}
\affiliation{Jo\v{z}ef Stefan Institute, Jamova c. 39, 1000 Ljubljana, Slovenia}
\affiliation{Faculty of Mathematics and Physics, University of Ljubljana, Jadranska ulica 19, 1000 Ljubljana, Slovenia}
\author{Peter J. Baker}
\affiliation{ISIS facility, Rutherford Appleton Laboratory, Chilton, Didcot, OX11 0QX, Oxfordshire, UK}
\author{Andrey Prokofiev}
\affiliation{Institute of Solid State Physics, Vienna University of Technology, Wiedner Hauptstrasse 8-10, 1040 Vienna, Austria}
\author{Matej Pregelj}
\email{matej.pregelj@ijs.si}
\affiliation{Jo\v{z}ef Stefan Institute, Jamova c. 39, 1000 Ljubljana, Slovenia}
\affiliation{Faculty of Mathematics and Physics, University of Ljubljana, Jadranska ulica 19, 1000 Ljubljana, Slovenia}

\date{\today}

\begin{abstract}

Fractional spin excitations in low-dimensional and frustrated magnets are commonly associated with sporadic spin fluctuations, as observed in quantum spin liquids, spin ices, and other complex magnetic states.
In this study, we investigate the spin dynamics in the amplitude-modulated magnetic ground state of \ch{AgCuVO4}, a spin-1/2 chain compound with competing intrachain interactions, by muon spectroscopy.
Our data reveal a Gaussian decay of muon polarization that remains remarkably resilient against applied magnetic fields---a behavior often interpreted as evidence of emergent sporadic spin dynamics.
However, density functional theory calculations of muon stopping sites and their quadrupolar level-crossing resonance spectra suggest an alternative mechanism: a two-state model where the muon randomly hops between two nearby sites, eventually equilibrating across both on a microsecond timescale via quantum delocalization.
These findings provide a microscopic explanation for the field-resilient muon polarization and indicate that such behavior may not always stem from intrinsic spin fluctuations, but can also arise from non-equilibrium muon dynamics.

\end{abstract}

\maketitle

\section{Introduction}

Unconventional magnetic ground states---such as quantum spin liquids, spin ices, and complex magnetic orders---are of significant interest both from a fundamental physics perspective and for potential technological applications \cite{lacroix2011introduction}.
These states often emerge due to reduced system dimensionality and/or competing magnetic interactions arising from geometrical frustration.
They exhibit a range of fascinating phenomena, including fractional spin excitations \cite{mourigal2013fractional,castelnovo2012spin}, and are predicted to host exotic phases such as multipolar magnetic orders \cite{hikihara2008vector,smerald2013theory}, for which a conclusive experimental confirmation is still pending.
Notably, emergent fractional excitations like spinons in spin-liquid states are considered promising candidates for future quantum computing technologies \cite{nayak2008non}.
A comprehensive understanding of the underlying physics is therefore highly desired.

A major challenge in studying these exotic states lies in their elusive nature and the difficulty of experimentally detecting their defining properties \cite{khatua2023experimental}.
A candidate observable is sporadic spin dynamics, driven by intrinsic fractional spinon excitations.
Muon spin spectroscopy ($\mu$SR) has emerged as a powerful technique for probing this phenomenon \cite{yaouanc2011muon,niedermayer2002muon}, owing to its exceptional sensitivity to local static and dynamic magnetic fields, and its ability to operate at temperatures as low as $\sim$20\,mK.
In particular, sporadic spin dynamics are often associated with a monotonic quasistatic (Gaussian) decay of the muon spin polarization, which is only weakly affected by an applied magnetic field, even when the applied field is significantly stronger than the average local fields one would infer from the muon polarization measured in zero field.
This resilience of the muon-polarization decay against external fields has been phenomenologically described by introducing a rather \textit{ad hoc} 
time (or field) rescaling factor linked to the exotic sporadic nature of the intrinsic spin fluctuations, assumed to cause strong local fields to be present at each muon site not persistently, but only a fraction of the time \cite{uemura1994spin,cho2020dynamic,dey2023local}.
However, a comprehensive theoretical explanation of this behavior is still lacking.
Despite this drawback, the model has been applied to a wide range of frustrated spin systems, including those on the kagome \cite{uemura1994spin,bono2004mu} and hyper-kagome lattices \cite{khuntia2016spin}, pyrochlores \cite{lago2005magnetic,xu2016spin,dey2023local}, double perovskites \cite{mustonen2022valence}, perovskite-like intermetallics \cite{tripathi2019quantum}, triangular lattices \cite{ortiz2023quantum}, and spin chain systems \cite{kojima1995muon}.

\begin{figure*}[!]
\centering
\includegraphics[width=\textwidth]{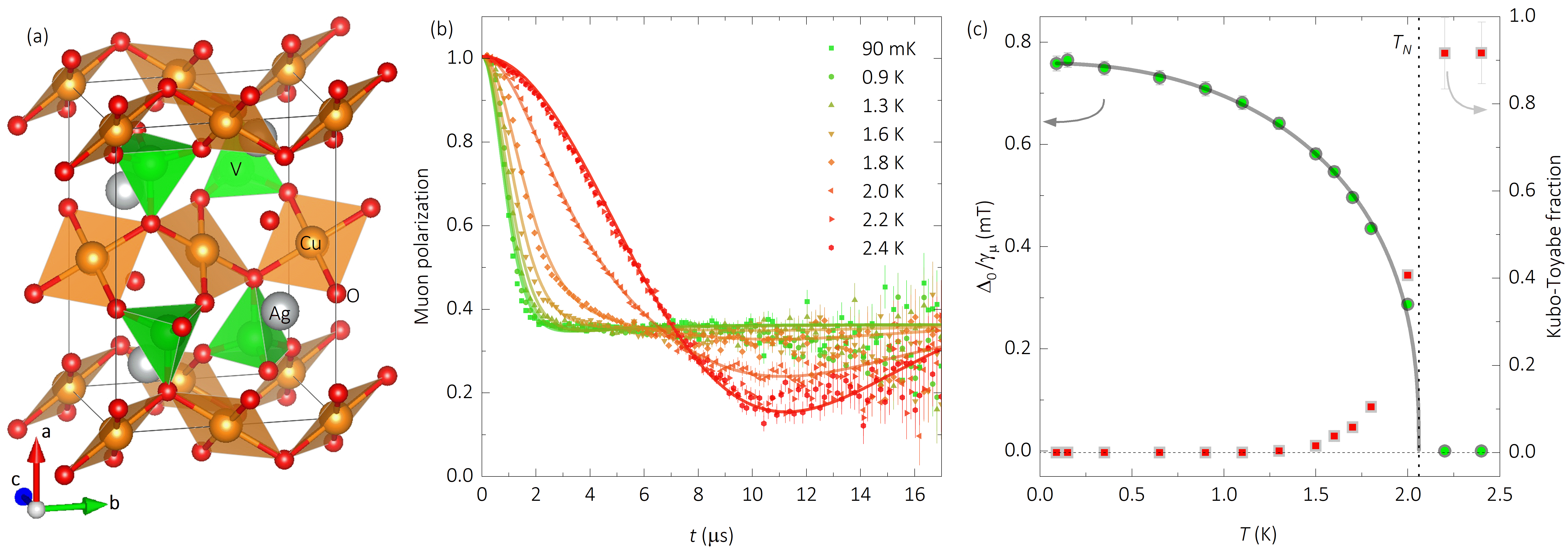}
\caption{
{\bf Temperature dependence of muon spin relaxation.}
(a) Crystal structure of \ch{AgCuVO4}.
(b) Muon polarization decay curves measured at various temperatures in zero applied field, along with the corresponding fits (see the main text for model details).
(c) Temperature dependence of the low-temperature mean local magnetic field and of the fraction of the nuclear Kubo-Toyabe contribution.
The solid line represents a fit to the Curie--Bloch expression $[1 - (T/T_N)^\alpha]^\beta$, while the dotted line marks the Néel transition temperature $T_N$.}
\label{fig-muSR-T-dep}
\end{figure*}

In this context, we highlight the quasi-one-dimensional $S = 1/2$ chain compound \ch{AgCuVO4} \cite{moller2009agcuvo}, in which the $\mu$SR signal in the magnetically ordered ground state exhibits a similar monotonic Gaussian decay of the muon polarization \cite{hromov2025incommensurate}.
This compound has an orthorhombic unit cell where magnetic Cu$^{2+}$ ions form spin chains along the $b$-axis, composed of corner-sharing [CuO$_4$]$^{6-}$ plaquettes \cite{moeller2008synthesis}. 
These chains are separated by nonmagnetic [VO$_4$]$^{3-}$ tetrahedra and Ag$^+$ ions [Fig.\,\ref{fig-muSR-T-dep}(a)]. The dominant intrachain interaction $J \approx 300$\,K is antiferromagnetic, while competing interchain interactions are estimated to be an order of magnitude weaker \cite{moller2009agcuvo}.
Below the Néel temperature $T_N = 2.0(1)$\,K, \ch{AgCuVO4} develops an incommensurate amplitude-modulated magnetic order, with collinear magnetic moments lying in the $ab$ plane \cite{hromov2025incommensurate}.
The orientation of these within the plane, however, remains undetermined.

Our present $\mu$SR study demonstrates that the magnetic moments align along the $b$ axis and, more significantly, reveals that the muon randomly hops between two nearby muon stopping sites and eventually equilibrates across both via quantum delocalization \cite{niedermayer2002muon,khatua2023experimental} on a microsecond timescale.
This dynamic behavior provides a natural explanation of the characteristic Gaussian decay of the muon polarization and its observed resilience to an applied magnetic fields, which does not involve exotic sporadic spin dynamics of the spin chains.
These findings offer an alternative perspective on the $\mu$SR response in frustrated spin systems, indicating that the reduced impact of the magnetic field may not always be due to sporadic spin dynamics, as commonly assumed.

\section{Results}

To investigate the Gaussian character of the $\mu$SR signal reported in our previous study~\cite{hromov2025incommensurate}, we performed extended $\mu$SR measurements on a powder sample of \ch{AgCuVO4} using the MuSR instrument at the ISIS Neutron and Muon Source in the UK.
These measurements effectively doubled the time window of our earlier experiments, extending our sensitivity to even smaller magnetic fields and relaxation rates.

At 2.4\,K, above the magnetic transition temperature, the muon polarization exhibits a distinct minimum due to nuclear dipolar fields that was previously beyond the experimental time window.
This behavior is well described by the Kubo-Toyabe (KT) model~\cite{yaouanc2011muon,niedermayer2002muon} [Fig.
\ref{fig-muSR-T-dep}(b)], assuming a random orientation of local magnetic fields with a Gaussian distribution with a width of $\Delta_\text{KT}/\gamma_\mu = 0.18(1)$\,mT, where $\gamma_\mu = 2\pi \times 135.5$\,MHz/T is the gyromagnetic ratio of the muon.
This result corroborates our earlier determination of the magnitude of nuclear fields based solely on a field-decoupling experiment at early times~\cite{hromov2025incommensurate}.

\begin{figure*}[!]
\centering
\includegraphics[width=\textwidth]{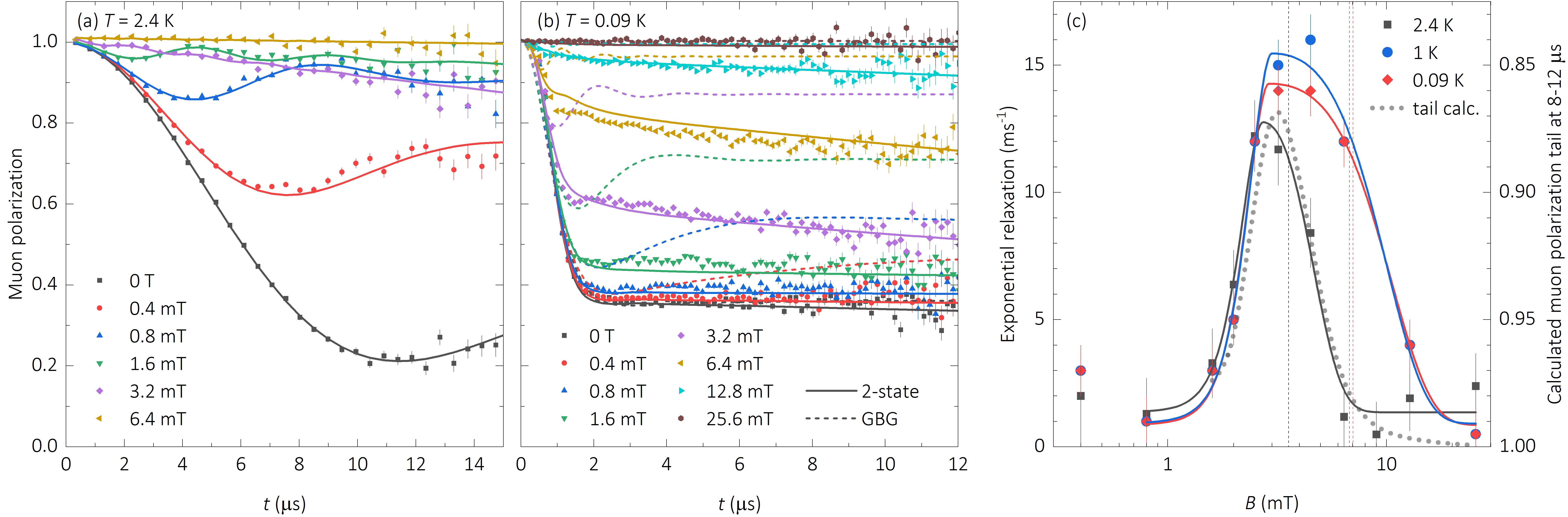}
\caption{
{\bf Muon spin relaxation in longitudinal magnetic fields.}
Muon polarization curves measured at (a) 2.4\,K and (b) 0.09\,K. In panel (a), the solid lines represent fits using the KT model for a longitudinal field, combined with exponential relaxation. In panel (b), solid and dashed lines correspond to fits using the two-state and GBG models, respectively. (c) Relaxation rates, $\lambda(B)$, extracted from the fits as a function of applied magnetic field at different temperatures. Solid lines are empirical fits to bigaussian peak shapes (see Methods), while vertical lines mark the positions of their corresponding mean values. Thick dotted line denotes the calculated QLCR effect on muon polarization averaged between 8 and 12 $\mu$s for muon stopping site \#2 (see text) in the absence of static fields from electrons, i.e., corresponding to the paramagnetic state.}
\label{fig-muSR-LF}
\end{figure*}

As the sample is cooled towards and below the magnetic transition at $T_N = 2.0$\,K, the muon polarization decay undergoes a distinct transformation [Fig.~\ref{fig-muSR-T-dep}(b)].
In particular, the minimum gradually becomes shallower, and at the base temperature of 90\,mK, the muon polarization curve evolves into a Gaussian-like shape with a nearly flat tail.
At this temperature, the muon decay is well described by the Gaussian-broadened-Gaussian (GBG) model [Fig.~\ref{fig-muSR-LF}(b)]~\cite{yaouanc2011muon}, in which the KT field distribution width $\Delta_\text{KT}$ is itself Gaussian distributed (modeling inhomogeneous random static fields) with a mean at $\Delta_0/\gamma_\mu = 0.75(5)$\,mT and a spread of $\Delta_{\text{GBG}}/\gamma_\mu = 0.59(5)$\,mT.
These parameters suggest a broad field distribution with substantial spectral weight near zero field, which is expected for incommensurate magnetic structures~\cite{yaouanc2011muon} that are proposed for the ground state of \ch{AgCuVO4}~\cite{hromov2025incommensurate}.

The temperature dependence of the muon polarization can, therefore, be modeled as a sum of a GBG function, representing the low-temperature electron magnetism, and a KT function, accounting for the nuclear fields
\begin{align}
P(t) =\ & (1 - f) \cdot P_\text{GBG}(\Delta_0, \Delta_\text{GBG}, t) \notag \\
& + f \cdot P_\text{KT}(\Delta_\text{KT}, t). 
\label{GBG+KT}
\end{align}
In this model, the high-temperature parameter $\Delta_\text{KT}$ (determined at 2.4\,K), and the ratio $\Delta_0/\Delta_\text{GBG} = 1.27(7)$ (determined at base temperature) are held fixed.
With only $\Delta_0$ and the fraction $f$ as free parameters, excellent agreement with experimental data is obtained [Fig.\,\ref{fig-muSR-T-dep}(b)].
Below 2\,K, the extracted magnetic field $\Delta_0/\gamma_\mu$ increases rapidly and saturates below approximately 0.5\,K [Fig.~\ref{fig-muSR-T-dep}(c)], exhibiting a typical order-parameter-like behavior
that is well described by the Curie--Bloch expression, $[1 - (T/T_N)^\alpha]^\beta$ \cite{marfoua2023ultra}.
A fit between base temperature and 2\,K yields values that are consistent with our previous study \cite{hromov2025incommensurate}.
The derived parameters are: the ordering temperature $T_N$\,=\,2.06(5)\,K, the critical exponent $\beta$\,=\,0.39(6), which is consistent with 3D magnetic order \cite{pelissetto2002critical,chaikin1995principles}, and the exponent $\alpha = 2.2(2)$, which is close to the expected $\alpha = 5/2$ for one-dimensional half-integer spin systems \cite{kobler2002temperature}.
These results indicate the formation of a broad distribution of static magnetic fields in the ordered phase, which fully replaces the paramagnetic phase at low temperatures, as evidenced by the KT fraction $f$ [Fig.\,\ref{fig-muSR-T-dep}(c)].

To probe local magnetic fields and spin dynamics in more detail, we performed longitudinal-field (LF) decoupling $\mu$SR experiments, applying magnetic fields along the initial muon polarization direction.
At 2.4\,K, above $T_N$, the muon polarization is well described by a nuclear KT function modified by the applied LF~\cite{yaouanc2011muon} and multiplied by a field-dependent exponential relaxation term [Fig.~\ref{fig-muSR-LF}(a)].
In a global fit across all fields (up to 6.4\,mT) we used shared parameters for the instrumental background and $\Delta_\text{KT}$, while fitting the applied field and exponential relaxation individually.
The extracted $\Delta_\text{KT}/\gamma_\mu = 0.185(5)$\,mT agrees with the zero-field value.

The derived exponential relaxation is negligible at low fields, increases sharply above 1\,mT, peaks at 13(2)\,ms$^{-1}$ near 3\,mT, and decreases at higher fields [Fig.~\ref{fig-muSR-LF}(c)].
This low-field maximum in relaxation is typically attributed to quadrupolar level-crossing resonance (QLCR) that occurs when the muon's Zeeman energy splitting matches the energy level gaps of a nearby quadrupolar (spin $I \geq 1$) nucleus.
For example, in a $\mu$SR study of a Cu single crystal, a relaxation peak was observed at $\sim$8\,mT with a maximum of 35(1)\,ms$^{-1}$~\cite{kreitzman1986longitudinal}, 
while in a $\mu$SR study of the kagome metal \ch{RbV3Sb5} vanadium QLCR resonances were found between 4--11\,mT~\cite{bonfa2025unveiling},
both comparable to our findings for \ch{AgCuVO4}.
We note that the resonant field depends on the electric field gradient at the nuclear site, which varies between different environments.

Measurements at base temperature in longitudinal fields up to 25.6\,mT offer additional insight into the magnetic ground state.
The calculated LF response of the derived GBG model clearly overestimates the influence of the applied field [dashed lines in Fig.~\ref{fig-muSR-LF}(b)]. 
Such behavior is commonly observed in frustrated magnetic systems~\cite{uemura1994spin,kojima1995muon,bono2004mu,lago2005magnetic,xu2016spin,dey2023local,mustonen2022valence,tripathi2019quantum,ortiz2023quantum,cho2020dynamic,khuntia2016spin} and is typically modeled by a field-rescaling factor \cite{uemura1994spin,cho2020dynamic,dey2023local}.
The latter is typically attributed to free spinon excitations in a spin-liquid state transiently altering the local field at the muon site~\cite{uemura1994spin}, but other intrinsic sporadic processes have also been proposed~\cite{dey2023local}.
In our case, applying a scaling factor of $\sim$1/3 reproduces the monotonic muon-polarization decay at longer times, but preserves pronounced oscillatory features at elevated fields (suppl. mat.)---as in the unmodified GBG model---that are clearly inconsistent with experimental observations.

We thus propose an alternative explanation for the apparent suppression of the applied longitudinal magnetic field based on a two-state (2-state) model~\cite{yaouanc2011muon}.
In this framework, the muon transitions between two distinct magnetic environments, a behavior typically attributed to diffusion between crystallographic sites.
The dynamics are modeled using the strong-collision approach, characterized by a well-defined transition rate $r$ between the two states.
In fact, the zero-field $\mu$SR signal in \ch{AgCuVO4} at base temperature is well described by an even simpler two-state model [Fig.\,\ref{fig-muSR-LF}(b)], assuming that the muon is initially in a non-relaxing state with constant polarization but eventually transitions to a final state characterized by a static KT field distribution~\cite{yaouanc2011muon,petzinger1980zero}.
By incorporating the effect of the longitudinal magnetic field on the KT distribution and including an additional slow exponential relaxation in the final state, we can accurately reproduce the full field dependence of the $\mu$SR signal [solid lines in Fig.\,\ref{fig-muSR-LF}(b)].
The best-fit parameters include a two-state transition rate $r = 1.1(1)$\,$\mu s^{-1}$ and a final KT field width $\Delta_\text{KT}^\text{2-state}/\gamma_\mu = 2.0(1)$\,mT, both assumed to be field-independent, while the relaxation rate $\lambda$ was allowed to vary with field.
The field dependence of $\lambda$ closely matches that observed in the paramagnetic state at 2.4\,K [Fig.\,\ref{fig-muSR-LF}(c)], supporting the interpretation that QLCR is the dominant relaxation mechanism in an applied magnetic field.
The extracted $\Delta_\text{KT}^\text{2-state}$ significantly exceeds the nuclear dipolar field scale $\Delta_\text{KT}$, which is consistent with the presence of long-range magnetic order of electronic spins.
Moreover, the mean value of the QCLR $\lambda(B)$ distribution [dotted lines in Fig.\,\ref{fig-muSR-LF}(c)] is shifted up by $\sim$3\,mT to higher fields at $T < T_N$, compared to the QCLR $\lambda(B)$ distribution in the paramagnetic regime [Fig.\,\ref{fig-muSR-LF}(c)].
This increase is consistent with the experimentally determined average field strength in the KT model~\cite{yaouanc2011muon} $\sqrt{2}\Delta_\text{KT}^\text{2-state}/\gamma_\mu = 2.8(2)$\,mT, resulting from the appearance of long-range magnetic order below $T_N$.
Finally, the inferred mean transition time $1/r = 0.9(1)$\,$\mu$s aligns with typical muon residence times---the timescale over which the muon experiences a change in its local magnetic environment reported in several studies~\cite{cox1987implanted,niedermayer2002muon}.

\section{Discussion}

\begin{figure}[!]
\centering
\includegraphics[width=\columnwidth]{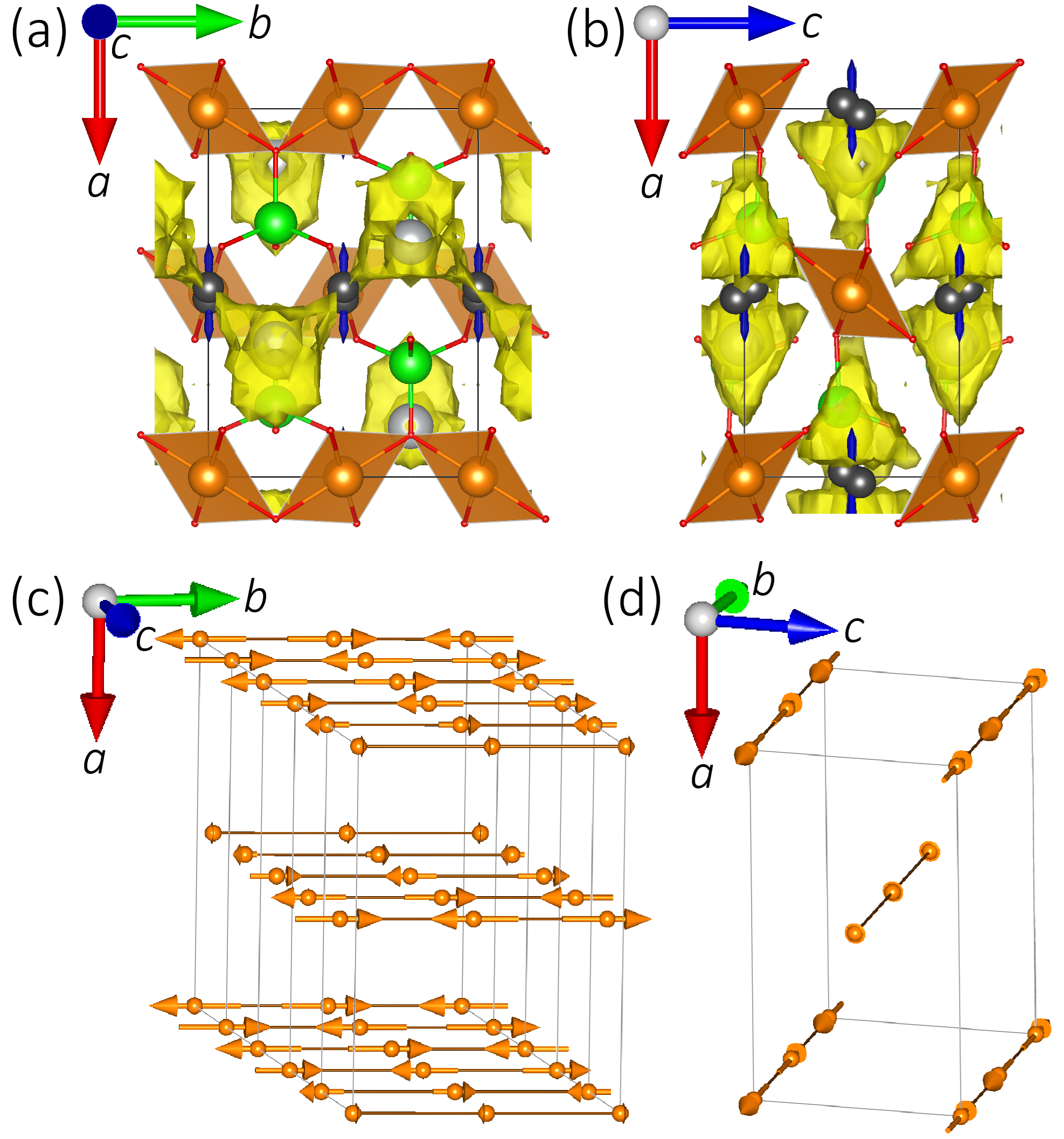}
\caption{
{\bf Crystal and magnetic structure of \ch{AgCuVO4} viewed along two different directions.} Panels (a) and (b) display the crystal structure, where atoms are represented as spheres: Cu (brown), V (green), Ag (light gray), and O (red). The muon stopping site is indicated by a dark gray sphere. Regions where the mean nuclear magnetic field equals 0.35\,mT are highlighted in light yellow, while areas corresponding to a magnetic field of 2.8\,mT (matching the experimentally determined value at base temperature), associated with the magnetic structure model with magnetic moments aligned along $b$, are shown in dark blue. Panels (c) and (d) display the corresponding magnetic order.}
\label{fig-struc-stop}
\end{figure}

To understand the physical origin of the two-state muon response, it is essential to identify the muon stopping sites in \ch{AgCuVO4}.
We performed density functional theory (DFT) calculations, which revealed five 
candidate sites (Table\,\ref{table_muon_sites}), with classical energy differences between them smaller than the expected quantum-muon zero-point energy (ZPE) correction, which is typically $\sim$0.4--0.7\,eV~\cite{blundell2021muon,gomilsek2023many,hotz2026anharmonic}.

To determine the most probable site, we calculated the local magnetic fields arising from nuclear dipole moments, which dominate the muon response in the paramagnetic phase~\cite{hromov2025incommensurate}.
The mean field value, computed for randomly oriented nuclear moments, corresponds to $\sqrt{2}\Delta_\text{KT}/\gamma_\mu$ in the KT model~\cite{yaouanc2011muon}.
The lowest calculated value is 0.24\,mT, which closely matches the experimentally determined value of 0.26(1)\,mT at 2.4\,K.
The region where nuclear fields are small, i.e., below  0.35\,mT [Fig.\,\ref{fig-struc-stop}(a),(b)], overlaps with three candidate muon stopping sites (\#2, \#4, and \#5).
Among these, one location (\#2)---at the edge of the 0.35-mT area and shown in Fig.\,\ref{fig-struc-stop}(a),(b)---exhibits a much lower classical-muon DFT energy (by $\sim$0.2\,eV) and a much more extended potential minimum (see Methods), which should give it a much lower additional quantum-muon ZPE~\cite{hotz2026anharmonic}), compared to the others, suggesting it as the most likely muon stopping site.

This is further supported by calculations of the QLCR spectrum for the DFT-derived muon stopping sites.
The calculations were performed using the electric field gradients, determined for the muonated structures via DFT, and taking into account the nearest quadrupolar nuclei, i.e., $^{51}$V and $^{63,65}$Cu, respectively.
Our results show that the QLCR spectrum is dominated by $^{51}$V nuclei at the measured fields with strong $^{51}$V resonances in the range 2--6\,mT (Supplementary Fig. S1; $^{63,65}$Cu resonances are found to only appear at fields ${\gtrsim}110$\,mT in this material), while the best agreement between the experiment and calculations is found for the muon stopping sites \#2 [Table\,\ref{table_muon_sites}, Supplementary Fig. S1, and Fig.\,\ref{fig-muSR-LF}(c)], corroborating the preceding analysis.
To the best of our knowledge, this makes \ch{AgCuVO4} only the second system to date, besides the kagome metal \ch{RbV3Sb5}~\cite{bonfa2025unveiling,pratt2026practical}, where a vanadium QLCR signal has been resolved.

To complete the local-field analysis, we also calculated dipolar magnetic fields for previously proposed incommensurate amplitude-modulated collinear magnetic structures, with moments aligned along either the $a$ or $b$ axis \cite{hromov2025incommensurate}.
The configuration with moments along $b$ yields an average field of $\sim$4(1)\,mT near the identified muon site, consistent with the experimental $\sqrt{2}\Delta_\text{KT}^\text{2-state}/\gamma_\mu = 2.8(2)$\,mT.
In contrast, the $a$-axis configuration produces comparable fields only in regions far from any DFT predicted muon site.
These results thus support a magnetic structure with magnetic moments aligned along the $b$ axis [Fig.\,\ref{fig-struc-stop}(c),(d)].

The muon response in the magnetically ordered phase provides further insight into muon spin dynamics. The two-state model, which accurately reproduces the longitudinal-field decoupling of the muon polarization, is compatible with several physical scenarios:

\begin{enumerate}
    \item \textbf{Muon diffusion between two physical sites:} The muon may diffuse between two crystallographically distinct sites, one experiencing rapid magnetic-field fluctuations and the other a static KT field distribution.
    However, this scenario is inconsistent with the paramagnetic response at 2.4\,K, where the longitudinal-field decoupling is well described by a single-site KT model.

    \item \textbf{Delayed magnetic ordering at the muon site:} The muon may reach its final stopping site immediately, initially experiencing fast magnetic-field fluctuations due to a local disruption of magnetic order.
    Over a timescale of $1/r \sim 1$\,$\mu$s, the system could relax into a static magnetic configuration.
    However, this explanation is unlikely, as the restoration of magnetic order should typically occur on much shorter timescales (10$^{-14}$ to 10$^{-10}$\,s) governed by phonon or magnon excitations~\cite{blundell2021muon}.

    \item \textbf{Muon hopping between nearby sites:} The muon may initially hop between two nearby \#2 sites, separated by just $\sim$0.5\,\AA~ [Fig.\,\ref{fig-struc-stop}], that are related by the local inversion center at the Wyckoff position $(0, 0, 1/2)$.
    The latter would thus represent a transition state TS between the two symmetry-related \#2 sites imposing a (classical-muon) energy barrier of $\delta E\sim$0.07\,eV, given by the energy difference between the TS and \#2 site (Table\,\ref{table_muon_sites}), which is an order of magnitude smaller than muon zero-point energy.
    Nuclear-field calculations show that the local magnetic field at these two \#2 sites on average differs only by $\sim$0.06\,mT in size and by $\sim$20$^\circ$ in orientation.
    Given a hopping rate of $\sim$10\,GHz, determined as $\delta E/h$, where $h$ is the Planck constant, the muon in the paramagnetic phase at \#2 site would effectively sample an averaged static field $B_{\text{avg}}\approx(0.35 - 0.06/2) \text{mT} = 0.32$\,mT, obtained as the mean value at the \#2 site reduced by one half of the average deviation, which is in good agreement with the experimentally determined $\sqrt{2}\Delta_\text{KT}/\gamma_\mu = 0.26(1)$\,mT at 2.4\,K.  

    Below the magnetic transition, the electronic magnetic fields at these two \#2 sites differ by $\sim$1.8\,mT in magnitude and are nearly antiparallel (deviate by $\sim$170$^\circ$). Consequently, rapid muon hopping would average out the electronic fields, effectively mimicking a non-relaxing muon state. Upon thermalization, the muon would occupy a quantum state delocalized over both sites (as the classical barrier height is small compared to the expected quantum ZPE of the muon), experiencing a broad KT-like field distribution with a mean value consistent with the experimental $\sqrt{2}\Delta_\text{KT}^\text{2-state}/\gamma_\mu = 2.8(2)$\,mT.
\end{enumerate}

Among these scenarios, the third provides the most consistent explanation of the experimental observations, without invoking unlikely physical mechanisms.

\section{Conclusions}

Using muon spin spectroscopy and density functional theory calculations, we have identified the most probable magnetic ground state of \ch{AgCuVO4}.
This state consists of antiferromagnetic spin chains with collinear magnetic moments aligned along the chain direction $b$, with their amplitudes incommensurably modulated perpendicular to the chains.
Additionally, the muon response exhibits a remarkable resilience to the applied magnetic field, which can be explained by a two-state model.
This model assumes that, on a microsecond timescale, the muon transitions from a fast-fluctuating regime to a fully static state.
We interpret this behavior as indicative of muon hopping between two nearby positions, followed by thermalization and quantum delocalization across both sites, resulting in a static magnetic field distribution.
This mechanism provides an alternative to the commonly proposed sporadic relaxation processes, which have been invoked to explain similar field suppression effects in frustrated spin systems \cite{ortiz2023quantum,dey2023local,cho2020dynamic,lago2005magnetic,kojima1995muon}, but which often lack a clear microscopic justification.

\section{Methods}

\subsection{Synthesis}
The powder sample was prepared by solid state reaction of AgVO$_3$ and CuO.
The powder mixture was re-ground and annealed at 395$^\circ$C several times.
Minor crystalline impurity phases CuO [0.6(1)\%] and AgVO$_3$ [1.4(1)\%] were identified in the final product  by neutron diffraction~\cite{hromov2025incommensurate}

\subsection{MuSR}
Muon spin relaxation experiments were performed on the MuSR instrument at the Rutherford Appleton Laboratory, United Kingdom, equipped with dilution refrigerator.
An almost completely flat background contribution to the $\mu$SR signal was estimated to $\sim$4(2)\% and was considered in all fitting procedures.
The measurements were performed in zero-field as well as longitudinal magnetic fields up to 256\,mT in temperature range between 0.9 and 2.4\,K.
Zero-field measurements were performed with longitudinal muon-spin polarization, aligned with beam axis.

The bigaussian-peak shape used to fit the field-dependent muon relaxation $\lambda(B)$ is a modified Gaussian line with different linewidth on the left and on the right side of the peak maximum, i.e., $\sigma_L$ and $\sigma_R$, respectively. 
The corresponding distribution has a mean value $x_{\text{mean}} = x_\text{peak} + \sqrt{\frac{2}{\pi}}(\sigma_R -\sigma_L)$, where $x_\text{peak}$ denotes the position of the peak maximum.

\subsection{DFT calculations}
The \mbox{MuFinder} program~\cite{huddart2022mufinder} was used to determine and analyze classical (point-particle) muon sites in \ch{AgCuVO4} via \textit{ab initio} DFT calculations. 
The latter were performed using the \mbox{CASTEP} plane-wave DFT code~\cite{clark2005first} with ultrasoft pseudopotentials and the PBE exchange--correlation functional~\cite{perdew1996generalized} with additional Hubbard repulsion~\cite{anisimov1997first}. 
An effective on-site Hubbard repulsions of $U_\mathrm{eff} = U - J_H = \SI{6.0}{\electronvolt}$, where $U$ is the bare Hubbard repulsion and $J_H$ is Hund's coupling, was chosen for \ch{Cu^{2+}} $3d$ orbitals throughout, consistently with typical Hubbard repulsion strengths found in related magnetic compounds with a similar local \ch{O^{2-}} environment around \ch{Cu^{2+}}~\cite{arh2020origin,iqbal2015paramagnetism,jeschke2015barlowite}. 

Muon-site DFT calculations were carried out on a $1 \times 2 \times 2$ supercell of \ch{AgCuVO4} to avoid finite-size effects around the implanted muon, while a $\SI{1400}{\electronvolt}$ plane-wave energy cutoff and $3 \times 2 \times 2$ Monkhorst--Pack grid~\cite{monkhorst1976special} reciprocal-space sampling was chosen to achieve numerical convergence. 
One positive elementary charge per supercell was applied to account for the positive charge of the muon. 
All calculations were converged to within a total energy tolerance of $\SI{0.1}{\nano\electronvolt\per\mathrm{atom}}$ in the self-consistent field (SCF) DFT loop, while geometry relaxation tasks were converged to within a tight $\SI{2}{\milli\electronvolt\per\angstrom}$ force tolerance on both the muon and the nuclei. 

To determine the classical muon sites in \ch{AgCuVO4} a random distribution of 52 symmetry-distinct initial muon positions with minimum muon--muon and muon--atom distances of $0.8$ and $\SI{1.0}{\angstrom}$, respectively, was chosen~\cite{huddart2022mufinder}. 
After DFT geometry relaxation from these initial positions, 5 stable candidate muon sites were identified using the connected-components symmetry-adapted clustering algorithm~\cite{huddart2022mufinder,gomilsek2023many} with a $\SI{1.0}{\angstrom}$ clustering tolerance (see Table\,\ref{table_muon_sites}).

The fact that the Wyckoff position $(0, 0, 1/2)$, which we denote TS, is in fact the saddle point, i.e., the point with the lowest energy barrier for transitions between the two symmetry-related copies of muon site \#2, was also checked via DFT.
This was performed by first geometry relaxing the atomic positions with the muon held in place by symmetry in position TS (with a net zero force) inside a $1 \times 2 \times 2$ supercell, as for other muon-site calculations, and then calculating the DFT $\Gamma$-point phonon normal modes and frequencies of the final structure~\cite{gomilsek2023many}. 
As expected, the normal mode related almost exclusively to a muon moving in the direction between the two symmetry-related copies of muon site \#2 had a very strong negative frequency of $\SI{-2051}{\cm^{-1}}$, while all other normal modes were positive-frequency. 
This proves that TS is indeed a classical transition state between the two copies of muon site \#2.

\begin{ruledtabular}
\begin{table}[!htb]
\caption{%
{\bf \textit{Ab initio} muon sites in \ch{AgCuVO4}.} 
Positions are given in unit-cell fractional coordinates, (classical-muon) energies are given relative to the lowest-energy muon site, 
and the reduced $\chi^2$/DOF refers to a global fit of all experimental LF decoupling data at 2.4\,K. 
The experimentally realized site \#2 is highlighted in bold. 
The classical transition state between two symmetry-related copies of this site is denoted by TS.
}
\begin{center}
\renewcommand{\arraystretch}{1.2}
\begin{tabular}{@{} l r r r @{}}
Site & Position & Energy (eV) & $\chi^2$/DOF \\
\hline
\#1 & $(0.89, 0.37, 0.80)$ & 0 & 30.4 \\
\textbf{\#2} & $\mathbf{(-0.02, 0.01, 0.46)}$ & \textbf{0.148} & \textbf{4.82} \\
\#3 & $(0.23, 0.06, 0.73)$ & 0.217 & 8.61 \\
\#4 & $(0.05, 0.33, 0.14)$ & 0.380 & 5.50 \\
\#5 & $(0.16, 0.64, 0.98)$ & 0.563 & 7.35 \\
\hline
TS & $(0, 0, 1/2)$ & 0.218 & 6.38
\end{tabular}
\end{center}
\label{table_muon_sites}
\end{table}
\end{ruledtabular}

\subsection{QLCR spectra calculations and fits}
To find which muon site is actually realized, we calculated the expected nuclear QLCR spectra for each of the candidate muon sites and compared them with experimental LF decoupling data at 2.4\,K (Supplementary Fig. S1). 
The QLCR spectra calculations were performed using a \mbox{CalcALC}-inspired Python program~\cite{calcalc}, by the same procedure as in Ref.~\cite{gomilsek2023many}, by taking into account the nearest two $^{51}$V nuclei (spin $I = 7/2$) and the two nearest $^{63,65}$Cu nuclei (both $I = 3/2$) around each muon site to achieve numerical convergence, and using the electric field gradients (EFG) calculated for the muonated structures via DFT. 
The measured muon asymmetry was then fit using the model
\begin{align}
A(t, B) &= (A_0 - A_\mathrm{bgd}) P_Q(t, B) e^{-\lambda_e(B) t} \\
&+ A_\mathrm{bgd} P_\mathrm{bgd}(t, B) ,
\end{align}
where $A_0$ is the initial asymmetry, $P_Q(t, B)$ is the QLCR signal at time $t$ and applied field $B$, $A_\mathrm{bgd}$ is the background contribution due to muons that stopped outside the sample, with a slowly decaying polarization function due to internal nuclear fields $P_\mathrm{bgd}(t, 0) = e^{-(\sigma_\mathrm{bgd} t)^2}$ and $P_\mathrm{bgd}(t, B) \approx 1$ for $B > 0$. 
Here, we took $P_Q(t, B) \approx P_Q^\mathrm{DFT}(f_\mathrm{cal}^{-1} t, B)$, where $P_Q^\mathrm{DFT}(t, B)$ is the QLCR signal calculated using raw EFG tensors from DFT, and $f_\mathrm{cal} \approx 1$ is a fitted calibration factor for how much DFT on-average overestimates the true EFG strength (see Ref.~\cite{gomilsek2023many}). 
The relaxation rate $\lambda$ due to the spin dynamics of electrons in the paramagnetic regime was described by the single-correlation-time, Redfield model
\begin{equation}
\lambda_e(B) = \frac{2 \Delta_\mathrm{PM}^2 \nu}{(\gamma_\mu B)^2 + \nu^2} ,
\end{equation}
where $\Delta_\mathrm{PM}/\gamma_\mu$ is the width of the local fluctuating field distribution and $\nu$ is their fluctuation rate. 
The reduced $\chi^2$/DOF of these fits are shown in Table\,\ref{table_muon_sites}, demonstrating that site \#2 fits the experimental data best. 
The results of the fits for muon site \#2 are shown in Supplementary Fig. 1. 
There, the EFG calibration factor $f_\mathrm{cal} = 1.039(4)$ (comparable to $f_\mathrm{cal} = 1.07(1)$ found in solid nitrogen, $\alpha$--\ch{N2}~\cite{gomilsek2023many}), the fluctuating field distribution width $\Delta_\mathrm{PM}/\gamma_\mu = 0.069(2)$\,mT and fluctuation rate $\nu = 0.65(4)$\,MHz, and the zero-field background relaxation rate $\sigma_\mathrm{bgd} = 68(1)$\,ms$^{-1}$.

\begin{acknowledgments}
This work has been funded by the Slovenian Research Agency (ARIS), projects No. J2-2513, J1-50012, N1-0356, J1-50008, N1-0345, and J2-60034, and  program No. P1-0125.
Computing resources were provided by a local cluster of the F5 department at the Jožef Stefan Institute (co-funded by ARIS), and by the STFC Scientific Computing Department’s SCARF cluster.

\end{acknowledgments}

\bibliography{Ref-AgCuVO4}

\newpage


\renewcommand{\thefigure}{S\arabic{figure}}
\renewcommand{\thetable}{S\arabic{table}}
\renewcommand{\theequation}{S\arabic{equation}}
\renewcommand{\thepage}{S\arabic{page}}
\setcounter{figure}{0}
\setcounter{table}{0}
\setcounter{equation}{0}
\setcounter{section}{0}
\setcounter{page}{1} 

\onecolumngrid

\begin{center}
\section*{Supplementary Materials for \scititle}

Anton Hromov$^{1,2}$, Andrej Zorko$^{1,2}$, Matja\v{z} Gomil\v{s}ek$^{1,2}$, Peter J. Baker$^{3}$, Andrey Prokofiev$^{4}$, Matej Pregelj$^{1,2}$\\
\textit{$^1$Jo\v{z}ef Stefan Institute, Jamova c. 39, 1000 Ljubljana, Slovenia\\
$^2$Faculty of Mathematics and Physics, University of Ljubljana, Jadranska ulica 19, 1000 Ljubljana, Slovenia\\
$^3$ISIS facility, Rutherford Appleton Laboratory, Chilton, Didcot, OX11 0QX, Oxfordshire, UK\\
$^4$Institute of Solid State Physics, Vienna University of Technology, Wiedner Hauptstrasse 8-10, 1040 Vienna}\\

\end{center}

\twocolumngrid




%
\begin{figure}[!]
\centering
\includegraphics[width=\columnwidth]{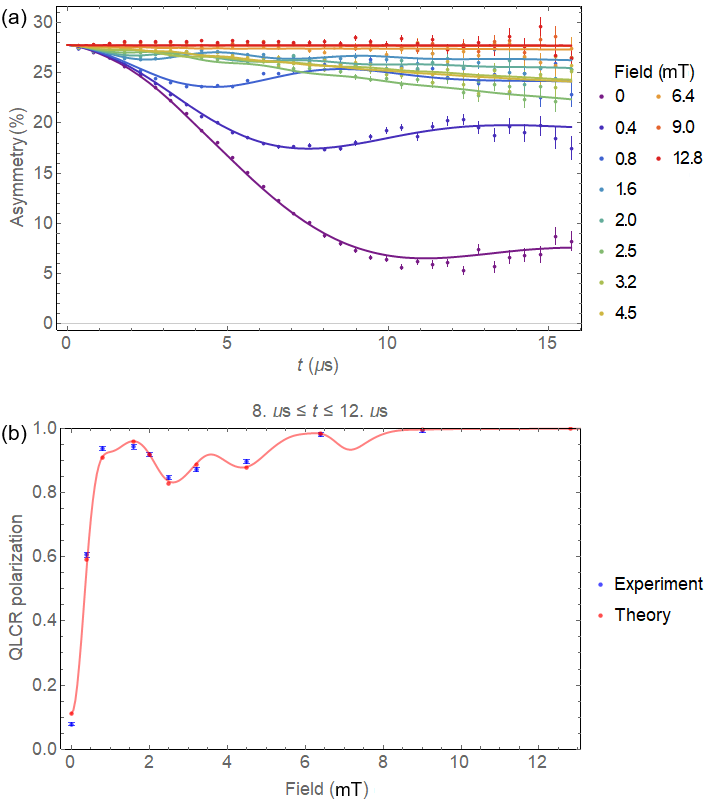}
\caption{
{\bf Calculated effect of QLCR on the muon polarization in \ch{AgCuVO4}.} Panel (a) show muon polarization in sveral longitudinal magnetic fields (symbols) and corresponding calculated response considering QLCR effect (solid lines) that are in almost perfect agreement. To emphasize the field dependence of the relaxation, we plot in panel (b) the average muon polarization tail at long times, i.e., in the range between 8 and 12\,$\mu$s, derived from the experiment as well as from calculations. The tails first rapidly increase, due to decoupling of the weak static nuclear magnetic fields, but exhibit a pronounced dip around 3\,mT due to QLCR.}
\label{fig-QLCR}
\end{figure}
$\\ \\ \\ \\ \\ \\ \\ \\ \\ \\ \\ \\ \\ \\ \\ \\ $

\begin{figure}[!]
\centering
\includegraphics[width=\columnwidth]{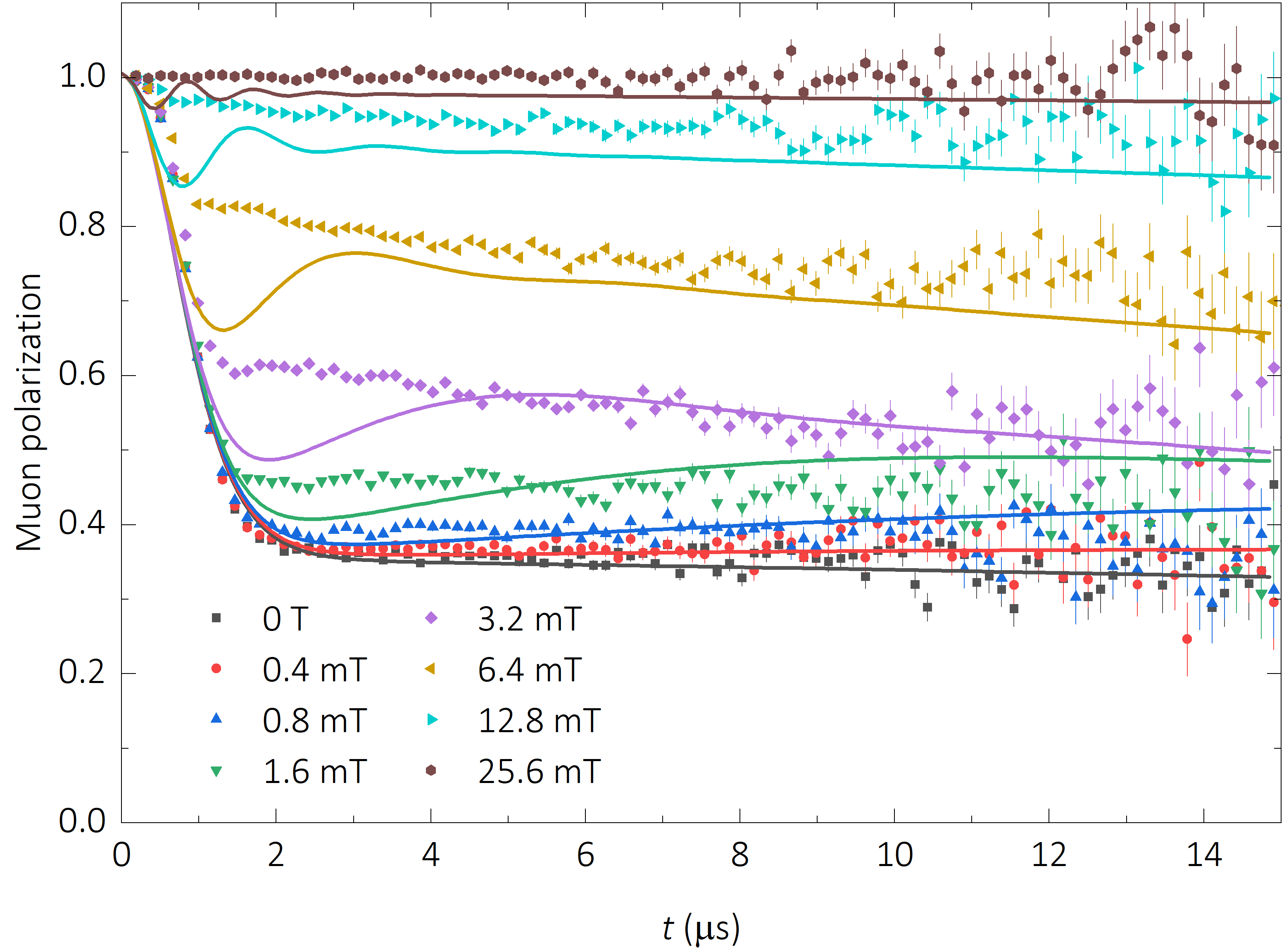}
\caption{
{\bf Longitudinal field decoupling of the muon polarization in \ch{AgCuVO4} at 0.09\,K.} Solid lines correspond to fits using the GBG model with a field-scaling factor of 1/3, $\Delta_0/\gamma_\mu$\,=\,0.8\,mT and $\Delta_{\text{GBG}}/\Delta_0$\,=\,0.81. While the model captures the muon polarization decay at longer times, it predicts pronounced oscillations at elevated magnetic fields and thus clearly deviates from the experimental results.}
\label{fig-struc-GBG-0p33}
\end{figure}

\end{document}